# GAME OF TONES: FACULTY DETECTION OF GPT-4 GENERATED CONTENT IN UNIVERSITY ASSESSMENTS

Mike Perkins [1]*, Jasper Roe [2], Darius Postma [3], James McGaughran [1], Don Hickerson [1],

[1] British University Vietnam, Vietnam.
[2] James Cook University Singapore, Singapore.
* Corresponding Author: Mike.p@buv.edu.vn

May 29, 2023

## Abstract

This study explores the robustness of university assessments against the use of Open AI's Generative Pre-Trained Transformer 4 (GPT-4) generated content and evaluates the ability of academic staff to detect its use when supported by the Turnitin Artificial Intelligence (AI) detection tool. The research involved twenty-two GPT-4 generated submissions being created and included in the assessment process to be marked by fifteen different faculty members.

The study reveals that although the detection tool identified 91% of the experimental submissions as containing some AI-generated content, the total detected content was only 54.8%. This suggests that the use of adversarial techniques regarding prompt engineering is an effective method in evading AI detection tools and highlights that improvements to AI detection software are needed. Using the Turnitin AI detect tool, faculty reported 54.5% of the experimental submissions to the academic misconduct process, suggesting the need for increased awareness and training into these tools. Genuine submissions received a mean score of 54.4, whereas AI-generated content scored 52.3, indicating the comparable performance of GPT-4 in real-life situations.

Recommendations include adjusting assessment strategies to make them more resistant to the use of AI tools, using AI-inclusive assessment where possible, and providing comprehensive training programs for faculty and students. This research contributes to understanding the relationship between AI-generated content and academic assessment, urging further investigation to preserve academic integrity.

## Introduction

**Background**

Artificial Intelligence (AI) has seen an upsurge in its application in various domains, and offers a wealth of opportunity for society, for example, helping medical professionals to diagnose diseases more quickly, and enhancing communication across borders (Pichai, 2023). However, AI's ability to seamlessly enable the production and reorganization of information carries with it associated risks. Among academics, existential conversations about the future of higher education and assessment are ongoing as teachers and students begin to acquaint themselves with the emerging challenges of a post-AI education landscape (Perkins, 2023). Such risks are poised only to grow, as it is estimated that AI applications are doubling in complexity and processing power every six months (Pichai, 2023).

One example of AI development creating risk in the academy can be seen in the recent public deployment of Large Language Models (LLM), such as GPT-3 and GPT-4, which have led to a 'wave' of attention not only from the public but also among policymakers and scholars (Bowman, 2023)). LLM applications such as ChatGPT, which is based on OpenAI's GPT models are an example of these generative-AI (Gen-AI) tools which exhibit sophisticated language generation capabilities. There are multiple benefits from an educational perspective to the use of LLMs, as they can effectively multiply the abilities of the user (Foltynek et al., 2023) in achieving their tasks. On the other hand, the sophistication of these tools in producing text that is fluent, detailed, and highly natural in tone means that distinguishing between human-written and AI-generated content is becoming increasingly difficult for faculty members to detect (Abd-Elaal et al., 2022; Köbis & Mossink, 2021). This leaves the opportunity for both unintentional misuse and intentional abuse of such tools to submit assessed work that has been partially or fully authored by AI, despite the availability of a growing range of software aimed at detecting AI generated text.

While dishonesty or misrepresentation of authorship is not a recent phenomenon, the risk of such cases ballooning unchecked carries potentially grave societal consequences. Bretag (2019) invites us to consider the implications of students who cheat their way to becoming doctors, engineers, and social workers by submitting work that is not their own, while Guerrero-Dib et al. (2020) highlight that intentional violations of academic integrity behaviours can continue into the workplace after students complete their studies. Thus, the risk must be treated as serious, and scholarship is required to identify potential solutions.

To this end, the European Network of Academic Integrity (ENAI) has released guidelines on the acceptability and use of AI tools in education, highlighting that if AI tools are used to produce assessed work which results in the awarding of credit or progression, then this could constitute academic misconduct. However, recognising that it is given that AI tools will likely form an important part of students' professional life in future, it is necessary for them to make use of these in an educative context and gain familiarity with the potential opportunities for AI (Foltynek et al., 2023).

## Study Objectives

The primary objective of this study is to explore the implications of AI-generated content in academic settings, with a particular focus on the detection and evaluation of such content. specific objectives can be detailed as follows:

1. Assess the quality and perceived authenticity of academic papers generated by AI tools, specifically the GPT-4 model.
2. Evaluate the effectiveness of AI detection tools, particularly Turnitin's AI detection tool, in identifying AI-generated content.
3. Assessing academic staff member's ability to use this information to make judgement on whether academic dishonesty may have occurred.
4. Understand the perceptions of academic staff regarding AI-generated content. This involves analysing the marks and feedback provided by academic staff on the AI-generated papers and identifying common themes and patterns.
5. Explore the implications of AI-generated content for academic integrity, specifically for potential adjustments to assessment strategies used by Higher Education Institutions (HEIs).

## Significance of the study

Our research contributes valuable insights to the expanding body of knowledge in several critical areas. Initially, we concentrate on the evaluation of the quality and perceived authenticity of academic papers generated by AI tools, specifically the GPT-4 model. This helps to gauge the potential impact of these tools on the academic landscape.

Secondly, we assess the effectiveness of AI detection tools, particularly Turnitin's AI detection tool, in identifying AI-generated content. This insight is crucial in the era of AI, as it informs our understanding of AI's potential to support faculty in the detection of work that may have been generated by AI tools. Thirdly, our study explores the ability of academic staff members to use the information provided by these detection tools during the marking process to make judgments regarding possible instances of academic dishonesty. This investigation is vital in maintaining the integrity of the academic process in the Gen-AI era. Fourthly, we aim to understand the perceptions of academic staff regarding AI-generated content. This involves analysing the marks and feedback provided by academic staff on the AI-generated papers and identifying common themes and patterns. Finally, we explore the implications of AI-generated content for academic integrity, specifically for potential adjustments to assessment strategies used by HEIs. This insight may help educators and HEIs design assessment strategies and tasks that accurately reflect students' achievement of course learning outcomes, despite the increased availability and potential prevalence of these Gen-AI tools.

The body of scholarship around LLMs is notably immature (Bowman, 2023), but understanding the implications of these advanced AI tools in relation to academic integrity is crucial. As the field of AI continues to evolve, it is essential for educators to adapt their assessment strategies to maintain academic rigor. One way to achieve this is to recognize the concept of cognitive offloading, a process described by Risko and Gilbert (2016), in which individuals rely on external resources to reduce their cognitive load. Dawson (2020) has equally described how assessment tasks can be developed in a way that makes the use of any such cognitive offloading tools transparent, potentially allowing for their legitimate use within the educational context. A specific example of LLMs being used for cognitive offloading is discussed by Dawson (2022, as cited in Sparrow, 2022). However, such methods have yet to gain widespread appeal, and as a result, it is presently still essential to understand whether an assessment was written by the submitting student, or by a Gen-AI tool.

While this research is predicated on the detection of AI generated text, we do not take the position that an effective method for solving misrepresentation of authorship and academic misconduct is by adopting a 'detect-and-punish' approach is desirable. Academic dishonesty is not necessarily a case of poor moral fibre, but is a nuanced, complex, and highly contextual event which can even be symptomatic of personal-social stress or struggle (Roe, 2022). Consequently, we take the epistemological position that by detecting AI-generated text in assessments, educators can find better ways to educate and train students, understand the use of AI, and adapt their assessment methods to maintain rigor and robustness for the benefit of all. As we acknowledge the irreversible nature of Gen-AI's integration into our systems, we must design assessment methods that effectively gauge students' knowledge while permitting some level of cognitive offloading in the development of assessment strategies. The findings of this study will serve as a basis for potential adjustments in assessment methodologies and guidelines across HEIs, ensuring academic integrity in the face of evolving AI technologies.

Artificial Intelligence (AI) has seen an upsurge in its application in various domains, and offers a wealth of opportunity for society, for example, helping medical professionals to diagnose diseases more quickly, and enhancing communication across borders (Pichai, 2023). However, AI's ability to seamlessly enable the production and reorganization of information carries with it associated risks. Among academics, existential conversations about the future of higher education and assessment are ongoing as teachers and students begin to acquaint themselves with the emerging challenges of a post-AI education landscape (Perkins, 2023). Such risks are poised only to grow, as it is estimated that AI applications are doubling in complexity and processing power every six months (Pichai, 2023).

**Structure of the Paper**

This paper is structured into seven main sections: introduction, literature review, methodology, results, discussion, limitations and future research, and conclusion. The literature review provides an overview of existing research on AI-generated content and its detection. The methodology section details the experimental design, the challenges in generating suitable content, and ethical considerations, while the results section presents the key findings. The discussion, limitations, and future research sections offer interpretations, implications, and suggestions for further investigation. Finally, the conclusion summarizes the study's findings and contributions to the field of academic integrity and assessment design.

# Literature Review

## AI-generated content and its impact on academic integrity

Over the past decade, AI-generated content has emerged as a significant concern for academic institutions around the globe. With the increasing sophistication of large language models (LLMs) like OpenAI's GPT-3.5 and GPT-4, questions have been raised about their impact on academic integrity(Abd-Elaal et al., 2022; Köbis & Mossink, 2021). These models can generate convincing content that closely mirrors human-written text, presenting a significant challenge to maintaining academic standards and ensuring fair assessment. These AI models can generate convincing content that may be difficult for educators to distinguish from human-written work, which poses a significant challenge to maintaining academic standards and ensuring fair assessments.

This advancement in AI has sparked serious ethical concerns related to potential misrepresentation of authorship and authenticity of academic work (Cotton et al., 2023; Perkins, 2023; Rahman & Watanobe, 2023; Rodgers et al., 2023; Rudolph et al., 2023). While there are tools available designed to detect the presence of Gen-AI content, their effectiveness has limitations (Malinka et al., 2023; Rudolph et al., 2023; Uzun, 2023), especially when users deliberately try and evade detection methods (Sadasivan et al., 2023). As a result, students may generate essays, research papers, or other academic assignments without acknowledging the assistance of AI or properly attributing the sources used by AI model (Perkins, 2023; Strzelecki, 2023). Such actions undermine the core principles of academic integrity, erode the educational value of assignments, and compromise the credibility of academic institutions. The debate on whether ChatGPT is a 'bullshit spewer' (Rudolph et al., 2023) highlights another concern: LLMs propensity for providing inaccurate or false information as fact (Azaria & Mitchell, 2023). This can be an indicator of the usage of LLMs in the production of text (Perkins, 2023).

Given these concerns, it's essential for institutions to establish clear guidelines and policies regarding the use of LLMs in assessed work. Comprehensive guidelines should outline the appropriate use of AI-generated content, citation requirements, and limitations (Crawford et al., 2023; Sullivan et al., 2023). The importance of proper attribution and students' responsibilities to maintain the integrity of their work should be explicitly emphasized. With clear guidance, institutions can help students navigate potential academic compliance issues associated with AI-generated content and promote responsible practices (Okonkwo & Ade-Ibijola, 2021; Sohail et al., 2023; Strzelecki, 2023).

## AI-generated content and its detection

Assessing the ability of Higher Education Institutions (HEIs) to detect the use of Large Language Models (LLMs) in student work is crucial for maintaining academic integrity Research on AI-generated content and its detection has primarily focused on earlier versions of GPT software with no currently available examples of studies exploring the detection of GPT-4 generated content. Here we summarise the current situation regarding the detection of Gen-AI content.

In relation to GPT-2, Abd-Elaal et al. (2022) demonstrated the challenges faced by academic staff in identifying LLM-generated content and the potential benefits of training for improving detection. Results showed that participants could correctly identify LLM-generated samples 59.5% of the time, only slightly better than chance. Similarly, Köbis & Mossink (2021) and Gunser et al. (2021) found that LLM-generated poems were difficult for experts to differentiate from human-written work. Fröhling & Zubiaga (2021) introduced a low-cost detection model that accurately identified GPT-2 and GPT-3-generated text. However, they noted the ethical challenges of deploying such detectors, which may inadvertently discriminate against non-native English speakers.

Bidermann and Raff (2022) demonstrated that more advanced models like GPT-J can evade detection tools like MOSS (Measure of Software Similarity), suggesting that newer LLMs might be even harder to detect. Similarly, Gehrmann et al. (2019) proposed a tool called GLTR to improve LLM-generated text detection, increasing accuracy from 54% to 74%. However, this study used GPT-2 and student participants. Solaiman et al. (2019) and Ippolito et al. (2020) also presented tools for detecting machine-generated text, although their applicability to newer LLMs remains untested. Research on GPT-3 detection suggests that human detection abilities decline as LLM complexity increases. For example, Kumar et al. (2022) found that participants from diverse backgrounds struggled to discern GPT-3-generated text from human writing. While not focused solely on academia, the study underscores broader detection challenges. Clark et al. (2021) reported that non-expert evaluators identified GPT-2-generated text with 57.9% accuracy, but GPT-3-generated text with only 49.9% accuracy. Training using LLM-generated examples marginally improved detection rates, however, Liang et al. (2023) found that by prompting GPT-3.5 to self-edit, detection rates for machine-generated scientific abstracts dropped from 100% to 13%, across seven different detector tools.

Despite the lack of empirical research at present which explores the detection of GPT-4 generated content, there is an ever-growing number of different commercial AI detection tools available (Uzun, 2023), including from OpenAI (the developer of GPT-4 and ChatGPT), and Turnitin. GPTZero, a free to use tool, is the self-proclaimed most popular current AI text detector with over one million registered accounts (GPTZero, n.d.-b). They claim to be able to classify 99% of human-written articles correctly and 85% of AI written articles correctly from several LLMs, including both ChatGPT (GPT-4) and Bard (LaMDA) (GPTZero, n.d.-a). OpenAI's detector proclaims far more marginal success rates, claiming to correctly identify only 26% of AI written text with a 9% false-positive rate (OpenAI, 2023b). Other services make much bolder claims Originality.AI claim to be the most accurate available detector of ChatGPT generated content with 95.93% true positive detection rates, and 1.56% false positive rates (Originality.AI, 2023). Similarly, Turnitin's detection tool, which was released in April 2023, claims to have 98% confidence in the ability to detect AI generated content whilst retaining a false positive rate of less than 1% (Turnitin.com, 2023). Other services including Netus AI, and stealthwriter openly promoting themselves as being able to paraphrase Gen-AI content to evade detection by other services. Netus AI state '*As AI content generation evolves, we continue to stay ahead of the curve*' (Netus AI, n.d.). This suggests that alongside the development of detection tools, equally we will see the emphasis on evasion of these tools increase.

Current methods used to identify machine-generated text can be broadly categorized into three groups: (1) Identifying the presence of 'watermarked' content; (2) Statistical outlier detection methods which seek out irregularities in the text generated; and (3) Classifiers that have been trained to distinguish between text generated by a machine and text written by a human (Krishna et al., 2023). Whilst watermarking is being promoted as a present solution to increase detection capabilities (Kirchenbauer et al., 2023). Sadasivan et al. (2023) demonstrated that the application of paraphrasing tools on top of generative text production can essentially invalidate the use of detection software, leading to the conclusion that even with watermarked text, highly sophisticated detection software can be easily evaded. A study by Mitchell et al. (2023) determined that before using a paraphrasing tool, DetectGPT accurately identified 70.3% of model-generated sequences from GPT2-XL. However, after using a paraphrasing tool to manipulate the content, the detection dropped to only 4.6%. It is critical to note that this was achieved with nominal semantic alterations. Solaiman et al. (2019) and Ippolito et al. (2020) also presented tools for detecting machine-generated text, although their applicability to newer LLMs remains untested.

These studies have explored various detection methods and their effectiveness in identifying AI-generated content, and although some detection techniques show promise, the constant advancements in AI models necessitate ongoing investigation and adaptation of detection strategies. The limited success shown of both human participants and detection tools in accurately detecting Gen-AI content underscores the threat to academic integrity in HEIs, and the need for an increased understanding of how detection works in practice.

## Advances in AI LLMs: GPT-4

The release of the GPT-4 LLM in March 2023 marked a major step in the journey of AI models in coming closer to achieving a true Artificial General Intelligence (AGI) system which can in theory perform any task a human can (Bubeck et al., 2023) and is the underlying model behind ChatGPT Plus, a paid-for-service offered on a subscription basis by OpenAI, as well as Bing search (Microsoft, 2023).

As with previous versions of the GPT LLM, GPT-4 utilizes a transformer-based model with self-attention for natural language processing (NLP) (Vaswani et al., 2017), and autoregressive modelling for time-dependent data analysis (Brown et al., 2020). However, it can accept inputs up to 16 times larger than GPT-3.5, and can generate outputs of up to 24,000 words: more than eight times larger than GPT-3.5 (Koubaa, 2023) and is now able to recognise and interpret visual inputs such as videos and images. (Campello de Souza et al., 2023). It has also shown improvements in reducing so-called 'hallucinations' and inappropriate responses (OpenAI, 2023a), whilst demonstrating advanced expertise in completing standardised tests such as a simulated bar exam and Graduate Records Examination (OpenAI, 2023a) and standardised IQ tests (Campello de Souza et al., 2023).

The existing literature suggests a major challenge to maintaining academic integrity is the ongoing advancement in AI LLMs, given the inability to accurately differentiate between human and AI-generated content. While detection tools do offer a degree of promise, their effectiveness can vary considerably, particularly when the content has been significantly altered or paraphrased. Notably, the research presents a key gap: there are no studies evaluating the ability of academic staff to identify AI-generated content produced by modern LLMs-given the improvements in the GPT-4 LLM identified above, this means that GPT-4 based LLMs should be able to perform at an even higher level than earlier models when it comes to developing output that will evade detection by academic staff. Furthermore, no studies have ventured into examining the efficacy of AI detection tools under conditions that mirror real-world scenarios often encountered during student paper assessment. This gap is particularly problematic for academic integrity, as it questions the principle of fairness and equality inherent in assessment methods used in HEIs across the globe. This study aims to address this gap, exploring the areas yet untouched in the literature.

## Methodology

### Context

The present study was carried out at a private university located in Southeast Asia with approximately 2000 students comprising four Schools. The University follows a UK-based curriculum and has achieved international quality accreditation by the UK Quality Assurance Agency (QAA). The language of instruction is entirely in English, with most students being non-native English speakers (NNES) and therefore studying in a second language.

All assessments involved in the study were part of partner collaborative programs, meaning that students were studying on programmes that led to the award of a UK degree from the partner institution. Assessments from the School of Business were chosen for use in this study as this is the largest School in terms of the number of programme offerings, and therefore provides the largest range of assessments being completed at any one time. The choice of one School ensured consistency in the general type of assessments being created by the research team, and the standards of evaluation of the academic staff members grading the work. Blind marking is carried out as standard using Turnitin Feedback Studio, and all fifteen participants involved in the study were experienced in using this system to carry out the assessment of student work.

Using a larger number of markers allowed us to obtain a more comprehensive understanding of the effectiveness of Gen-AI content in evading detection by faculty members with diverse backgrounds and expertise. By including this wide range of markers, the study aimed to account for any potential biases or inconsistencies that might arise from a single marker's assessment style or approach.

### Ethical considerations

The study obtained approval from the University Registrar, the Dean, and the University Human Ethics Committee before commencement. The faculty members involved in the study were informed about the research and freely gave informed consent, as well as being offered the choice to opt-out at any time. This decision was made after extensive discussions within the research team about whether it would be possible to avoid informing academic staff members about the fake student submissions and utilise a blind study methodology. However, it was ultimately determined that, given the additional time requirements to mark these fake submissions, academic staff members needed to be given an option to opt out of the study. Students' privacy was protected by creating a separate profile for submitting AI-generated content, ensuring that no student profile was linked to these test submissions.

### Experimental design

To investigate the robustness of university assessments against GPT-4 generated content, this study employed the following methodology:

1. A list of all end-of-semester assessments for January 2023 in the School of Business (n=70) was reviewed to determine whether Gen-AI tools may potentially be used to provide responses.
2. Assessments such as presentations, demonstrations, and exams were excluded, and assessment papers for the remaining assignments were downloaded by the research team (n=25).
3. Using GPT-4 accessed through the ChatGPT Plus interface, the research team developed responses (n=22) to assessment papers that could be reasonably expected to be submitted by a student, employing techniques to create responses that met the requirements of the assessment paper and reduce the likelihood of detection by AI detection software.
4. One test student profile was created and enrolled in the final assessments of all modules where responses could be created. The papers developed by the research team were added to the list of submissions due for marking by the broader faculty team participating in the study.
5. The faculty team were informed of the presence of the falsified papers and were asked to flag any submission they suspected of being AI-generated.

6. Following the marking process, the final grade awarded by the faculty team, the scores from the Turnitin AI detector, whether the paper was reported for academic dishonesty, and other relevant information were recorded.

7. Debriefing was provided to all faculty members involved in the study to inform them of the AI-generated paper, and training was arranged for all faculty to support them in the identification of AI generated content in the future.

In the January 2023 semester, the School of Business conducted a total of 70 end-of-semester assessments. These assessments were selected as the population for this research. However, certain types of assessments—namely exams, presentations, and primary research projects were excluded from this research as these assessments could not be reproduced or would not result in an acceptable submission. Consequently, the final list was narrowed down to 22 submissions that could be adequately assessed and generated via GPT-4 accessed through ChatGPT Plus. The choice was made to use the GPT-4 model for the creation of the submissions, as no previous studies have tested the effectiveness of this model in generating academic content. GPT-4, being one of the most advanced LLMs available at the time of the study, provided a unique opportunity to assess the potential impact of cutting-edge AI technology on academic integrity.

The research team generated a variety of assessments for marking by the research participants. These assessments included essays, reports, and case analyses over a diverse spectrum of topics including e-commerce portfolios, development of personal development plans, marketing strategies, consumer behaviour reports, reflective pieces on work experience, and essays on supply chain management. These assessments were then submitted to the relevant classes for marking by the relevant members of faculty using Turnitin Feedback Studio with the AI detection feature enabled. Participants were provided with instructions on how to use this newly available feature in the grading process, and how to use this information in making judgement as to whether potential academic dishonesty had occurred. Any suspected cases of authorship misrepresentations, including the contract cheating, plagiarism, and the use of Gen-AI tools in the creation of work are reported to the academic compliance team who arrange for a standardised investigation into whether academic misconduct has occurred.

**Challenges faced in generating suitable content**

Standard essays or assignments that prompted students to choose a country, organization, or topic and complete specific tasks were relatively straightforward to generate by the research team. More challenging were tasks that required reflections on "lived experiences" such as internship experiences or personal development plans. These could be created through providing small amounts of additional detail when creating the prompts entered in ChatGPT, for example, providing information on student societies. This helped in the production of assessments that matched the context of the University and would be less likely to be identified as AI generated. Some assessment tasks proved more challenging. For example, tasks that required primary research, or the pre-approval of the use of selected databases, or topic choice before beginning the task, were unsuitable for creation using Gen-AI tools and could not be created. However, this does not mean that Gen-AI tools could not be used by students taking part in similar assessments, only that these cases were not suitable for inclusion in this study.

During the submission creation process the research team engaged in so called 'prompt engineering' techniques to both obtain responses that met assessment requirements, but also took advantage of various adversarial techniques to evade possible detection by academic staff and AI detectors, using a range of techniques These included requests to ChatGPT to produce output at a level of complexity or language comprehension more typically seen in NNES students, requests to add spelling or grammatical errors for authenticity, regulating output word count, and ensuring real academic references. (Marche, 2022; Perkins, 2023)Despite Marche's (2022) claims about the obsolescence of traditional essays due to Gen-AI tools, it proved challenging to align the output of AI-generated content with the required word and content limits that could feasibly have been produced by a NNES student, with repeated requests needing to be made to modify content as per the research team's specifications. However, the output provided by ChatGPT had a tendency to revert to its typical style, resisted incorporating deliberate errors, and continued to provide non-existent references - a key indicator of Gen-AI created content, according to Perkins (2023).

Although Sadasivan (2023) highlighted the use of paraphrasing tools as a key method to evade AI detection software, as one of the aims of the study was to explore the potential of GPT-4 generated content to evade detection, this step was not taken in order to retain only output created by GPT-4. Apart from basic formatting of the text (such as providing titles in bold) to create realistic looking submissions, this meant that the research team also chose to make no other manual adjustments to submissions to help them better fit the requirements and potentially achieve higher marks.

## Results

### Summary results

Table 1 shows the results of the experiment. This table shows a summary of the required task, the number of true submissions for the assessment, the mean score achieved by true students, the mean score of the test student, the percent above or below the mean score obtained by the test student, what percentage of the test assessment that was identified by Turnitin AI detected as being generated by AI tools, and whether or not the paper was reported through formal processes as a potential breach of academic conduct regulations. The papers are sorted in descending order by the percentage of AI generated content that was identified by the Turnitin AI detection tool.

| Case | Summary of required task | # of submissions | Mean Score | Test submission score | % above or below mean score | Test student Turnitin detected AI % | Paper Reported for Academic Misconduct? |
|---|---|---|---|---|---|---|---|
| 1 | Organization reputation & brand management essay. | 15 | 67 | 70 | +4% | 100% | No |
| 2 | Strengths personal reflection. | 66 | 46 | 33 | -28% | 100% | Yes |
| 3 | E-commerce portfolio: search engine optimization, databases, communications. | 67 | 51 | 35 | -31% | 85% | Yes |
| 4 | RACE framework digital marketing report. | 97 | 56 | 51 | -9% | 81% | Yes |
| 5 | Credit policy essay: improvement, efficiency, data, risk management. | 10 | 57 | 49 | -15% | 81% | Yes |
| 6 | Consumer behaviour analysis & strategy recommendations. | 5 | 52 | 75 | +44% | 80% | Yes |
| 7 | Organizational structure analysis & recruitment recommendations. | 66 | 36 | 38 | +5% | 79% | Yes |
| 8 | Accountancy changes due to online technology. | 5 | 57 | 50 | -12% | 78% | Yes |
| 9 | Global supply chain competitive advantage essay. | 57 | 54 | 65 | 19% | 78% | Yes |
| 10 | Innovation-based marketing plan. | 99 | 59 | 18 | -69% | 75% | No |
| 11 | Leadership strategies & change impact evaluation. | 27 | 55 | 68 | +24% | 70% | No |
| 12 | Local market PESTEL analysis & marketing strategies. | 7 | 70 | 50 | -29% | 48 | No |
| 13 | Personal Development Plan: SWOT analysis and SMART goals. | 66 | 46 | 56 | +23% | 46% | Yes |
| 14 | Local financial markets, debt collection & regression analysis essay. | 26 | 56 | 52 | -7% | 40% | No |

| Case | Summary of required task | # of submissions | Mean Score | Test submission score | % above or below mean score | Test student Turnitin detected AI % | Paper Reported for Academic Misconduct? |
|------|--------------------------|------------------|------------|-----------------------|------------------------------|-------------------------------------|------------------------------------------|
| **15** | Objective setting & budgeting business blog. | 10 | 61 | 65 | +6% | 30% | Yes |
| **16** | Work experience reflection & recommendations. | 6 | NA* | NA* | NA* | 18% | Yes |
| **17** | Digital marketing blogs: web presence, social media. | 44 | 48 | 49 | +3% | 18% | No |
| **18** | Career action plan with skills gap analysis. | 54 | 59 | 70 | +18% | 13% | No |
| **19** | Managerial work experience reflection evaluation. | 6 | NA* | NA* | NA* | 13% | Yes |
| **20** | Company reputation & brand analysis, improvement, communication strategies. | 51 | 60.4 | 53 | -12% | 4% | No |
| **21** | Agile business essay: service operations & targets. | 88 | 48 | 62 | +29% | 0% | No |
| **22** | Personal development plan: communication, self-management. | 91 | 50 | 37 | -26% | 0% | No |

*Table 1 Full results*

\* These two papers were identified as test student submissions by the markers and were not awarded a grade. They have been excluded from the calculation of the mean grade.

Table 2 shows a summary of these results:

| Value | Figure |
|-------|--------|
| Mean number of submissions | 44 |
| Mean score of true submissions | 54.4 |
| Mean score of test student | 52.3 |
| Mean % obtained by the test student above or below the true submission mean score | -4.9% |
| Mean % of AI content identified by Turnitin AI Detect tool | 54.8% |
| Mean % of papers highlighted as containing some AI-generated content by Turnitin AI Detect tool | 91% |
| % of papers reported by academic staff through the formal academic misconduct process. | 54.5% |

*Table 2 Summary of results*

The findings of this study shed light on the significant disparity in the capability of academic staff to identify AI-generated content. In total, 22 academic papers were submitted as part of the investigation, of which a little over half, 12 (54.5%), were identified as potentially AI-generated by the academic staff. The mean score for genuine submissions was found to be 54.4, whereas the AI-generated content received a mean score of 52.3, showing an average deviation of -4.9% from the true submissions. The mean percentage of AI content detected by Turnitin AI detect was 54.8%, with 91% of the submission being highlighted as containing some AI generated content and potentially raising the suspicions of markers. No significant differences were found between the average scores given to detected papers (51.7%) and undetected papers (52.6%). This result indicates that the detection of AI-generated content did not substantially impact the grading process.

The feedback provided by the academic staff members offers valuable insights into the quality of the AI-generated submissions. One key aspect that emerged is the differing perceived quality of the AI-generated papers. Some markers praised the quality of the work, with one stating that the submission has "*many ideas that might be worthwhile to pursue*" and noting that the paper was "*well supported with good research and good clear thinking*" (Case 1). On the other hand, other markers found the submission to be lacking in depth or focus, with one stating that the paper "*lacks in-depth research*" (Case 7) and another noting the "*unfocused*" nature of the paper (Case 21).

Certain characteristics of the AI-generated papers, such as writing style and structure, were noted in the feedback from academic staff. For example, one marker mentioned the "*confusing*" writing style (Case 12), while another noted the lack of proper referencing (Case 18). Similarly, another marker highlighted the "*lack of personality and visuals*" (Case 15). In other cases, markers criticised the paper for its "*very long introduction*" and "*problem with sources*" (Case 8), while another noted that the paper "*did not address the Learning Outcomes*" (Case 3). However, it is important to note that we cannot determine whether the presence of these specific features specifically influenced academic staff to make these comments, as it is plausible that the AI percentage score provided by Turnitin could have influenced their evaluations and guided their comments.

## Discussion

### Interpretation of results

The results of this study provide valuable insights into the robustness of university assessments against GPT-4generated content. Turnitin's ability to detect 91% of the generated submissions as containing some AI-generated content is promising, despite the deployment of adversarial techniques to evade detection by the research team. As it is likely that any detection will raise the suspicions of markers assessing the paper for potential academic misconduct violations, this shows that Turnitin AI detect may be a valuable tool in supporting academic integrity. However, as only 54.8% of the total amount of AI-generated content was detected by Turnitin, there is still significant room for improvement, as demonstrated by the challenges faced by academic staff in accurately detecting AI-generated papers, even with Turnitin's results available.

The similar average scores between detected and undetected papers imply that the detection of AI-generated content did not significantly influence grading. The lower grades obtained by submissions where the task had additional requirements highlights the role of assessment design in mitigating the potential impact of AI-generated content on grading outcomes. Incorporating assessment tasks that require the application of specific frameworks or the pre-approval of topics and data sets may serve as additional barriers to AI-generated submission. Assessment strategies involving students providing group submissions therefore appears to be a potential technique for improving the ability of academic staff to detect AI generated content.

It is important to emphasise the differences between the effectiveness of Turnitin in identifying AI-generated content, compared to the faculty's ability to use this information and make a judgement as to whether this constitutes an academic misconduct violation. Although Turnitin correctly identified 91% of the papers as containing AI-generated content, faculty members only formally reported 54.5% of the papers as potential cases of academic misconduct. This may have occurred due to an underestimation of the amount of AI-generated content detected by Turnitin in many cases: the mean percentage of content identified by the Turnitin AI detection tool was only 54.8%, despite 100% of the content of all papers being generated with AI tools. This is in direct contrast to claimed results of a maximum of 15% of AI written text being missed in the AI reports (Turnitin.com, n.d.) and indicates that the use of adversarial techniques regarding prompt engineering is an effective method in evading the AI detection tools. This disparity may have led faculty members to believe that cases with a small amount of AI detected content were either were false positives (as can sometimes be seen in Turnitin similarity detection tools) or that a small amount of AI tool usage was acceptable. For example, one paper received a Turnitin AI score of 13%, but the academic staff member still marked it as authentic, possibly due to the perceived relatively low percentage of AI-generated content. However, some papers with very high AI scores (for example, Cases 1, 10, and 11) were not reported as potential academic misconduct violations.

This highlights the need for additional training to be provided to faculty members to enable them not only to rely on the raw results of the Turnitin AI detection tool but also to recognize the identifiers of potential AI tool usage that were experienced by the research team in the development of submissions. Some of these identifiers include overly complex language, under word length papers, core-texts and content discussed in class not being covered, and falsified references. A further consideration is that shorter and non-continuous texts are more likely to return errors in detection (Turnitin.com, n.d.). Consequently, the use of non-continuous text (for example bullet points) may be an adversarial strategy that could be employed by well-informed students to circumvent AI detection tools. Adopting such an approach, combined with other interventions as mentioned above, severely limits the effectiveness of both AI tools and faculty observation to identify AI-generated text.

In three cases (15, 16, and 19), the marker not only reported the cases as potential academic misconduct offences but were able to correctly identify the test student submissions. The following feedback was provided for Case 15:

> '*Dear Research Team, the lack of personality and visuals was the giveaway for this paper. I had also given a company to each group. However, in a large cohort this could slip through the net*'.

This set of submissions had only 10 true submissions, suggesting that with a smaller set of assessment submissions, markers may be more aware of the specific circumstances of each student or group of students' writing capabilities and progress on assessment, and would be more able to detect the usage of AI tools in submitted work. As a result, it is likely that AI detection tools will be of less benefit in smaller cohorts in which the instructor has a greater deal of insight into student ability. On the other hand, larger cohorts which include hundreds of students and are perhaps taught by a team of teachers will lead to less familiarity with the cohort and expected output, meaning that detectors are likely of higher value. This also suggests that adjusting modes of assessment to include group submissions may be one potential option for increasing the likelihood of spotting any AI generated content.

In some cases, faculty members still provided relatively high marks to papers which had high AI scores and were correctly identified as being generated by AI, such as a paper with a Turnitin AI score of 80% receiving a mark of 75 out of 100 (Case 6). Under the UK grading system, this equates to a First-Class result-the highest possible classification of results. This suggests that AI-generated content may still be perceived as valuable or well-structured, potentially raising concerns about the reliability and integrity of assessments if AI detection tools are not being used.

A notable number of outliers with very low attained scores were observed, particularly in assessments that required the use of frameworks not specified in the assessment paper or the pre-approval of topics and data sets (Cases 2, 3 7, 10 and 22), even though in some of these cases there was no indication of AI tool usage by either Turnitin AI detect, or the individual faculty member. Closer explorations of the feedback provided in case 22 revealed that some academic staff members expected specific usage of tools from core-texts: '*You have not used any of the SWOT from your text book*', or that additional exemplars were provided to students in class and on the Learning Management system but not used in the test submission: '*This PDP is meant to be in paragraphs as demonstrated in the exemplar on Canvas*'. This suggests that even though Gen-AI tools available to students may support them in the development of work used for submission, without engagement with course materials or attendance at class, the final marks achieved may be very low.

The results of this study highlights the potential of AI detection tools in identifying AI-generated content, but also underscores the challenges faced by academic staff in detecting such content. The integration of Turnitin's AI detection tool into the standard marking interface of a service used by many HEIs globally, as well as the high overall accuracy rate of detecting that Gen-AI content is present in a submission, means that student papers created using Gen-AI tools which may have otherwise gone undetected can now be more easily identified. However, there remains a gap in both faculty members' ability to recognise and respond to such content as well as the accuracy of the software in being able to identify Gen-AI text which has been produced using adversarial methods. As the results obtained from the current experiment demonstrate, these tools are not infallible, and must be combined with appropriate training to support faculty in developing assessment which are robust to the usage of Gen-AI tools, as well as spotting the signs of these tools being used. Providing additional training and support to faculty members will help them develop a more comprehensive understanding of AI-generated content indicators and improve the overall integrity of the assessment process.

**Implications for AI detection software development**

The results of this study also have implications for the ongoing development of AI detection software. While Turnitin demonstrated a broadly accurate detection rate of Gen-AI content, continuous improvements and updates to AI detection algorithms will be necessary to keep pace with the evolving capabilities of AI models like GPT-4. Collaboration between academia and AI detection software developers will be essential to ensure the effectiveness of these tools in maintaining academic integrity.

However, it is important to recognize that this 'arms race' scenario between Gen-AI tools and AI detectors is not sustainable. Although Turnitin claim that their detector mostly retains its ability to identify Gen-AI content after paraphrasing (Turnitin.com, n.d.), Sadasivan et al. (2023) demonstrate that the accurate detection of AI-generated text by software can almost always be thwarted by adversarial methods, such as the use of Automated Paraphrasing Tools (APTs) or cross-translation into different languages. Students will inevitably find ways around any such tools used to detect AI output, so it is likely to be more effective to focus on the adjustment of assessment strategies to either discourage AI-generated submissions, or changing assessment strategies entirely to encourage their use, rather than relying solely on detection software.

Additionally, AI detection software can sometimes produce false positives, particularly when faced with work written by non-native English speakers (NNES) (Liang et al., 2023). It has been demonstrated that students who have lower levels of English proficiency are more likely to commit academic misconduct violations (Perkins et al., 2018), and Fröhling and Zubiaga (2021) highlight the ethical challenges of deploying AI detectors, which may potentially discriminate against EFL students by incorrectly identifying human-created text as machine-written. This is a particular concern in higher education institutions with a high concentration of non-native English-speaking students, as the use of AI detection software could inadvertently undermine the academic integrity of their work. In light of these challenges, it is crucial for higher education institutions to strike a balance between utilizing AI detection software and adapting assessment strategies to account for the evolving landscape of AI-generated content. By working in tandem with AI detection software developers and incorporating more robust assessment methods, higher education institutions can uphold academic integrity while recognizing the potential benefits and limitations of AI tools in the academic environment.

**Implications for assessment design and academic integrity**

The findings of this study have several implications for assessment design and academic integrity in higher education. First, incorporating assessment tasks that necessitate unique frameworks, pre-approved topics, or data sets can help reduce the likelihood of AI-generated submissions being successful. By designing assessments that are either more challenging for AI tools to complete, or where their use could be easily identified, institutions can encourage students to engage more deeply with the subject matter and rely less on AI-generated content.

A move more towards overall assessment strategies involving a significant amount of written or oral examinations, presentations, or invigilated assessments may also be an option that some HEIs may choose to use, although this strategy reduces the opportunity to demonstrate to students the potential benefits of Gen-AI tools after graduating. An alternative approach would be recognising that a certain amount of cognitive offloading with the use of Gen-AI tools could be considered acceptable for students. For example, institutions can increase the depth of detail required in assignments, knowing that students may rely on AI tools to help them generate content. This approach acknowledges the reality of AI tools' presence in the academic environment and could encourage students to use them in a responsible and productive manner (Kumar, 2023). Designing assessments that explicitly require the use of AI tools and include marks for discussing and critically evaluating the prompt engineering that occurs to generate effective output would take this one step further. As graduates will likely be expected to use these tools in their future careers, institutions should prepare students by training them in the responsible and effective use of AI tools. By incorporating AI tools into the assessment process, HEIs can ensure that students are ready to leverage the benefits of AI in their professional lives while maintaining a strong commitment to academic integrity. We recognised that many HEIs may not be willing or able to take such an approach more focused on the detection of Gen-AI content, Faculty members should be equipped with the skills to recognize the subtle signs of AI-generated content and respond appropriately to maintain the integrity of the assessment process (Kumar, 2023).

HEIs must also focus on providing training and resources to faculty members to improve their ability to identify AI-generated content, and train and support students to use these tools in an appropriate manner. As academic misconduct training has been identified as a successful approach in mitigating instances of academic misconduct (Perkins et al., 2020), this will foster a more comprehensive approach to academic integrity that is not solely reliant on AI detection tools.

**Limitations and Future Research**

While providing key insights, this study does have certain limitations. The research took place within a single Business School, which could limit how widely the findings can be applied to other fields or institutions. Moreover, the small group of academic staff who evaluated the AI-produced papers may not fully capture the range of possible responses to AI-generated content. The possibility of faculty reusing comments from comment banks, a common practice to save time and maintain consistency, also wasn't assessed in this study, and could have influenced the responses provided.

The novelty of the Turnitin AI detection software used could also have affected the results. This was the first time the academic staff had used this tool, and although all participants were used to marking with Turnitin, the AI score provided is a distinct concept from the commonly used similarity index where, unlike the AI index, a low level may be overlooked or considered as normal by markers. Such unfamiliarity could have potentially skewed their perception or evaluation of which papers were AI-generated, underlining the need for further training and familiarization with these novel tools in future studies. This study focused on content created by one LLM: GPT-4. As other LLM models such as Bard by Google, or LLaMA by Meta, are becoming more popular, future research would benefit from including these models to ensure a broader perspective. Finally, given the smaller scale of this study, we should be careful when extending these results to broader contexts.

Extending the scope of the study to include larger participant groups, broader ranges of assessment types, multiple institutions and disciplines would further enhance the generalisability of the findings. Additional research could also explore the impact of specific faculty training programs on the ability to detect AI-generated content, determining which strategies are most effective. As more advanced AI models continue to emerge, future studies should also evaluate the efficacy of evolving AI detection tools in identifying content generated by these sophisticated models.

## Conclusion

The rapid evolution of AI technologies and their adoption into academia raises crucial questions about their implications for academic integrity and assessment processes. This study, focusing on the capacity of Turnitin's AI detection software and academic faculty to identify AI-generated content, provides initial insights into this complex landscape.

While the findings demonstrate the potential of Turnitin's AI detect tool in supporting academic staff to detect AI-generated content, the relatively low detection accuracy of the participants underscores the need for further training and awareness. The lack of significant differences between the mean scores achieved by the AI-generated submissions and the student submissions further highlights the potential implications for the integrity of university assessments, especially if AI detection tools are not being utilised. Additionally, the existence of potential adversarial strategies for defeating AI detection, such as the use of paraphrasing tools and prompt engineering techniques to adjust the output of LLMs further emphasises the significance of this challenge. Although in this case, Turnitin AI detect was relatively robust to the use of these adversarial techniques, there will likely be a continued 'arms race' (Roe & Perkins, 2022) between the tools designed to produce and detect AI generated content.

The social dimension of AI and LLM generated text and detection should not be underestimated. An area that has not been touched on in this research is that of the viability and accessibility of detection tools in higher education in much of the world. While there has been explosive growth in the AI field, access to technology in a learning environment remains inequitable across socioeconomic divisions (Reimers et al., 2020). The disparity in digital access and AI exposure, often termed the 'digital divide', can lead to further educational inequalities (Cullen, 2001). Moreover, the financial burden of implementing AI detection, particularly in developing countries and underprivileged communities, can be cost-prohibitive. As Turnitin have made clear that the AI detect feature will require additional subscription costs from January 2024 (Turnitin.com, n.d.), this suggests that even when such tools are effective, there are limitations to their ease of implementation and scalability, and it is likely that HEIs with less access to resources will face the biggest risk of GPT-generated content in undermining academic integrity.

This research also raises questions regarding the ongoing development of AI detection software and its effectiveness. As advanced AI models and adversarial techniques continue to evolve, there is a pressing need for the constant improvement and updating of detection algorithms. Yet, it is important to acknowledge that detection software alone will not be sufficient in maintaining academic integrity. The future of academia should focus on striking a balance between harnessing AI technology responsibly and maintaining the sanctity of academic integrity. The development of assessment tasks that explicitly involve the use of AI tools, coupled with comprehensive training programmes for faculty and students, could shape the future of assessment in higher education. Not only will this promote the responsible use of AI tools, but it will also equip students with skills that will be invaluable in their future careers, all the while upholding the integrity of the assessment process.

While this study is an initial step towards understanding the interplay between AI-generated content and academic assessment, it has also opened the door to further research. As AI continues to evolve, so too should our understanding of its impact on academia. We encourage HEIs, academic staff members, and software developers to continue to engage with this area of research, contributing to the development of robust strategies that both utilise the potential of AI and preserve the principles of academic integrity. Ultimately, this study serves as a call to action for academic institutions and faculty to adapt to the rapidly changing landscape of AI technology. As AI continues to permeate various facets of our lives, it is crucial that the academic community remains abreast of these advancements and their implications for the integrity of our assessment processes and the overall quality of education.

## Statements and Declarations

**Funding**

No funding was received for conducting this study.

**Competing interests**

Mike Perkins, Darius Postma, James McGaughran, and Don Hickerson are currently employed by the University where the study took place. Jasper Roe was previously employed by the same University.

**Ethics approval and consent**

The study obtained approval from the institution's Human Ethics Committee before commencement and all participants gave informed consent with options to opt-out of the study at any point in time.

**Data availability**

The full commentaries left by academic staff members when marking the test submissions cannot be shared due to the confidentiality of assessment results and associated comments. All other data generated or analysed during this study are included in this published article.

**Author contributions**

Mike Perkins conceived and designed the study. Data collection and analysis were performed by Mike Perkins, Darius Postma, James McGaughran, and Don Hickerson. The first draft of the manuscript was written by Mike Perkins, with all authors subsequently revising the manuscript. All authors read and approved the final manuscript.

**LLM usage disclaimer**

The preparation of this manuscript was supported with the use of the GPT-4 LLM, accessed through ChatGPT. The authors take full responsibility for the accuracy of the content presented and the views and opinions expressed in this paper are solely those of the authors and do not represent the official stance of OpenAI.

**Acknowledgements**

The authors are very grateful for the support of the academic staff members who participated in the study, and for the ongoing support of the University Examinations Office team who provided technical assistance throughout the project.